# Training performance of $Nb_3Sn$ Rutherford cables in a channel with a wide range of impregnation materials

S. Otten, A. Kario, W.A.J. Wessel. J. Leferink, H.H.J. ten Kate, M. Daly, C. Hug, S. Sidorov, A. Brem, B. Auchmann, P. Studer, T. Tervoort

***Abstract*—Training of accelerator magnets is a costly and time consuming process. The number of training quenches must therefore be reduced to a minimum. We investigate training of impregnated $Nb_3Sn$ Rutherford cable in a small-scale experiment. The test involves a Rutherford cable impregnated in a meandering channel simulating the environment of a canted-cosine-theta (CCT) coil. The sample is powered using a transformer and the Lorentz force is generated by an externally applied magnetic field. The low material and helium consumption enable the test of a larger number of samples. In this article, we present training of samples impregnated with alumina-filled epoxy resins, a modified resin with paraffin-like mechanical properties, and a new tough resin in development at ETH Zürich. These new data are compared with previous results published earlier. Compared to samples with unfilled epoxy resin, those with alumina-filled epoxy show favorable training properties with higher initial quench currents and fewer training quenches before reaching 80% of the critical current.**

## I. Introduction

LENGTHY training remains an issue for epoxy resin impregnated $Nb_3Sn$ accelerator magnets [1]. A high number of training quenches before reaching nominal current is not considered acceptable for large-scale application due to the high cost. Although the precise cause of a training quench is hard to detect, there is a theory that it is related to strain energy from cool-down and Lorentz force, which is released on failure of the impregnant [2]. Possible failures include cracks within the resin volume and debonding between the resin and metal surfaces. To prevent formation of cracks, resins with high toughness are currently being developed [3][4][5]. Another improvement is to add fillers to the resin, which reduces the thermal expansion mismatch and prevents propagation of cracks.

In a collaboration of the Paul Scherrer Institute (PSI) and the University of Twente, we developed a small-scale training experiment for impregnated Rutherford cables called BOnding eXperiment (BOX) [6]. The experiment requires only 1 m of cable, and the sample is energized by a transformer which needs only 50 A for the primary coil. The relatively low cost of the experiment allows us to test a larger number of samples. The BOX experiment is complementary to the subscale CCT coils built at the Lawrence Berkeley National Laboratory (LBNL), which are more representative of a full-size magnet [7].

In our previous publication [8], we presented training curves of cables impregnated with different unfilled resins and paraffin wax. In this work, we present new results on BOX samples modified to reduce training by adding glass fiber or $Al_2O_3$ filler. A new tough resin developed at ETHZ is also tested [9].

## II. Experimental Method

### A. Sample preparation

The training curves are measured on $Nb_3Sn$ Rutherford cables made at LBNL [10]. The cables consist of 21 strands of Bruker OST RRP 108/127 wire with a diameter of 0.85 mm. The cables are insulated using a braid of S-2 glass of 0.075 mm thickness. More properties of the cable can be found in table I.

The cable is placed in an aluminum bronze or stainless steel holder with a meandering channel, which was sand blasted for a clean surface (see Figure 1). The dimensions of the channel are 2.5 mm x 10.8 mm. The cable is heat treated at 210 °C for 72 hour, 400 °C for 48 hour and 665 °C for 50 hour in an argon atmosphere. After heat treatment, voltage taps are placed in each bend using silver epoxy, and finally the sample is impregnated.

TABLE I
CABLE AND STRAND PROPERTIES

| Parameter | Value | Unit |
|---|---|---|
| Strand diameter | 0.85 | mm |
| Sub-element size | ≤55 | µm |
| Filament twist pitch | 19 ± 3 | mm |
| Cu/Sc ratio | 1.2 ± 0.1 | - |
| Sub-element configuration | RRP 108/127 | - |
| Number of strands | 21 | - |
| Keystone angle | 0° | - |
| Cable twist pitch | 70 | mm |
| Cable height without insulation | 1.475 | mm |
| Cable height with insulation | 1.785 | mm |
| Cable width without insulation | 9.85 | mm |
| Cable width with insulation | 10.16 | mm |

Manuscript receipt and acceptance dates will be inserted here. Acknowledgment of support is placed in this paragraph as well.

S. Otten, A. Kario, W. Wessel. J. Leferink, and H. ten Kate are with the University of Twente, Enschede, the Netherlands (e-mail: s.j.otten@utwente.nl).

M. Daly, C. Hug, S. Sidorov, A. Brem and B. Auchmann are with the Paul Scherrer Institute (PSI) Villigen, Switzerland. B. Auchmann is also with CERN, Geneva, Switzerland.

P. Studer and T. Tervoort are with ETH Zürich, Switzerland.

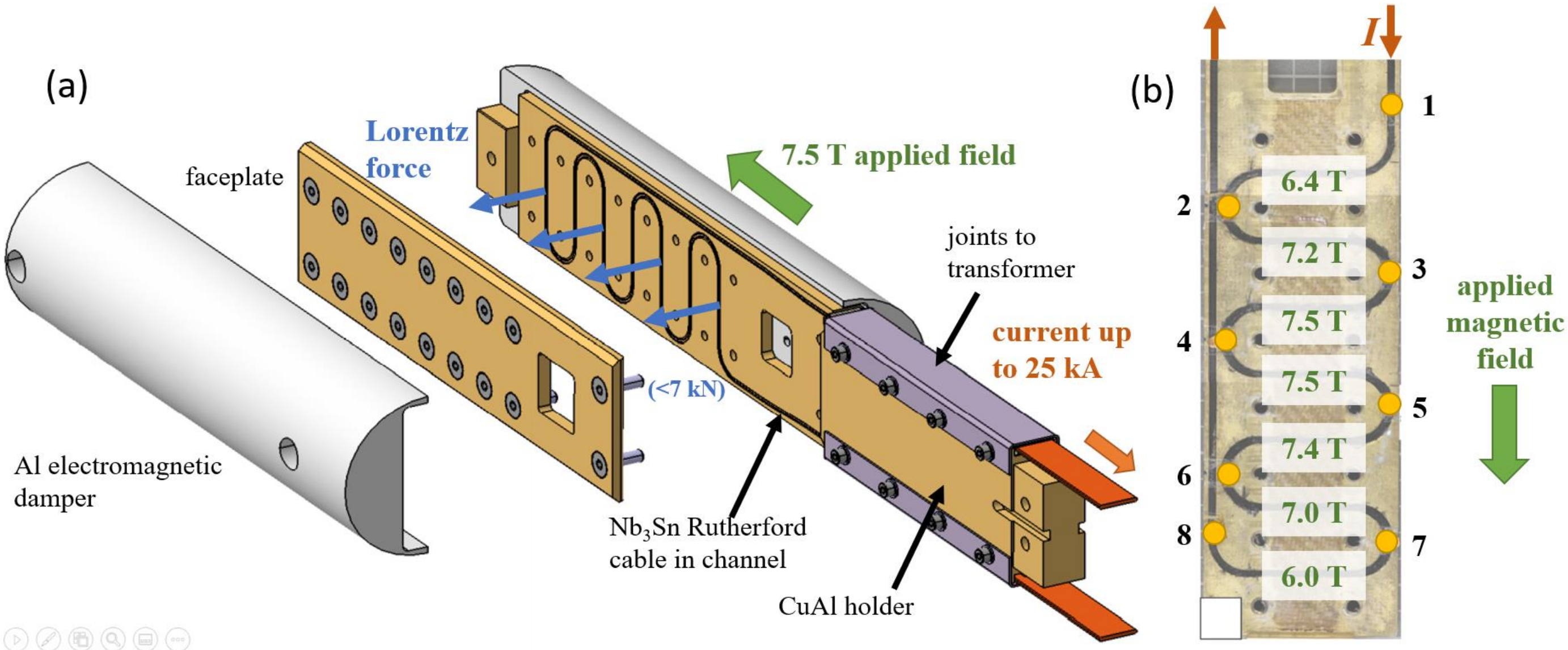

Fig. 1. (a) "BOX" sample holder for measurement of training curves. (b) Location of the voltage taps and applied magnetic field magnitude at each of the seven perpendicular segments.

The different samples presented in this paper are listed in Table II. There are five new samples with different resins and cable insulation in an attempt to reduce training:

- CTD-101K with $Al_2O_3$ filler: the channel was filled with alumina powder by sedimentation before vacuum impregnation with CTD-101K. The glass insulation was removed from this cable because it is incompatible with the filler. Instead, the holder has a ceramic coating for insulation (Aremco SGC4000);
- CTD-101K double sleeve: the cable was inserted into a second S-2 glass sleeve in order to increase the ratio of glass/epoxy;
- CTD-101K weakened: a non-stoichiometric mixture of the CTD101K resin components with an about 5 times lower fracture toughness than CTD101K;
- Stycast 2850FT: an alumina-filled resin cured at room temperature. The glass insulation was removed from this cable because it is incompatible with the filler. Instead, the holder has a ceramic coating for insulation (Aremco SGC4000);
- ETHZ Cryoset 2 M: an interim version of a new tough epoxy resin in development at the ETH Zürich [9].

TABLE II
BOX SAMPLES: IMPREGNATION AND CURING CONDITIONS

| Sample/impregnation | Cure | $T_g$, melting temperature |
|---|---|---|
| Paraffin (1)* | - | 52-56 °C |
| Paraffin (2)* | - | 52-56 °C |
| Stycast 2850FT/23LV | 24 h at 25 °C | 68 °C |
| NHMFL mix 61* | 16 h at 25 °C + 24 h at 102 °C | 73 °C |
| Araldite MY750* | 6 h at 40 °C + 3 h at 80 °C | 57 °C |
| CTD-101K (1)* | 5 h at 110 °C + 16 h at 125 °C | 145 °C |
| CTD-101K (2)* | 5 h at 110 °C + 16 h at 125 °C | 145 °C |
| CTD-101K with $Al_2O_3$ | 5 h at 110 °C + 16 h at 125 °C | 145 °C |
| CTD-101K double sleeve | 5 h at 110 °C + 16 h at 125 °C | 145 °C |
| CTD-101K weakened | 5 h at 110 °C + 16 h at 125 °C | |
| CTD-701X* | 1 h at 40 °C + 2 h linear ramp 40-120 °C + 1 h at 120 °C | 131 °C |
| ETHZ Cryoset 2 M | 2 h at 80 °C + 3 h at 150 °C | |

Samples with * are from our previous publication [8] and are included for comparison. Glass transition temperatures $T_g$ from [5] and [11] or datasheets.

### B. *Training curve and critical current measurement*

To generate the Lorentz force, a magnetic field of $B_a$ = 7.5 T is applied using a solenoid magnet. A magnetic field of 7.5 T was chosen because maximum Lorentz force of $B_a*I_c(B_a)$ peaks at this magnitude. The channel with the cable has seven 35-mm-long segments perpendicular to the applied magnetic field. At the expected critical current of 23.6 kA, these segments experience a Lorentz force of 6.2 kN parallel to the wide cable surface.

In order to provoke a training quench, the sample current is ramped up at 200 A/s until a quench occurs. The current as well as the voltage over each segment are recorded using a multichannel oscilloscope (Yokogawa DL850EV).

After training, the critical current is measured with a criterion of $E_c$ = 10 μV/m. The in-field sample length is not well defined due to the meandering shape. Therefore we assume a length of 40 mm and define the critical current at the lowest current at which the voltage in at least one segment reaches 0.4 μV.

## III. RESULT

### A. *Training curves*

In Figure 2, the training curve for the sample impregnated with Araldite MY750 is shown. There are two regimes that can be distinguished. The first quench occurs at 14.2 kA. The following quenches are at gradually higher currents until the critical current of 22.8 kA is reached at quench 20. These quenches start in unpredictable locations, which is indicative of training. After 25 quenches, most quenches start in the high-field region (segment 3-4) at 104% of the critical current. These quenches are most likely caused by heating due to the superconducting to normal transition. In Figure 3, the quench currents $I_q$ of different samples are plotted normalized to the critical current $I_c$, which can be found Table III. As already presented before [8], the paraffin-impregnated showed no training with all quenches starting at 102-104% of the critical current. The sample with Stycast 2850FT also showed decent training behavior with the first quench at 88%, higher than all other resin-impregnated samples. This sample reached the critical current at quench 3.

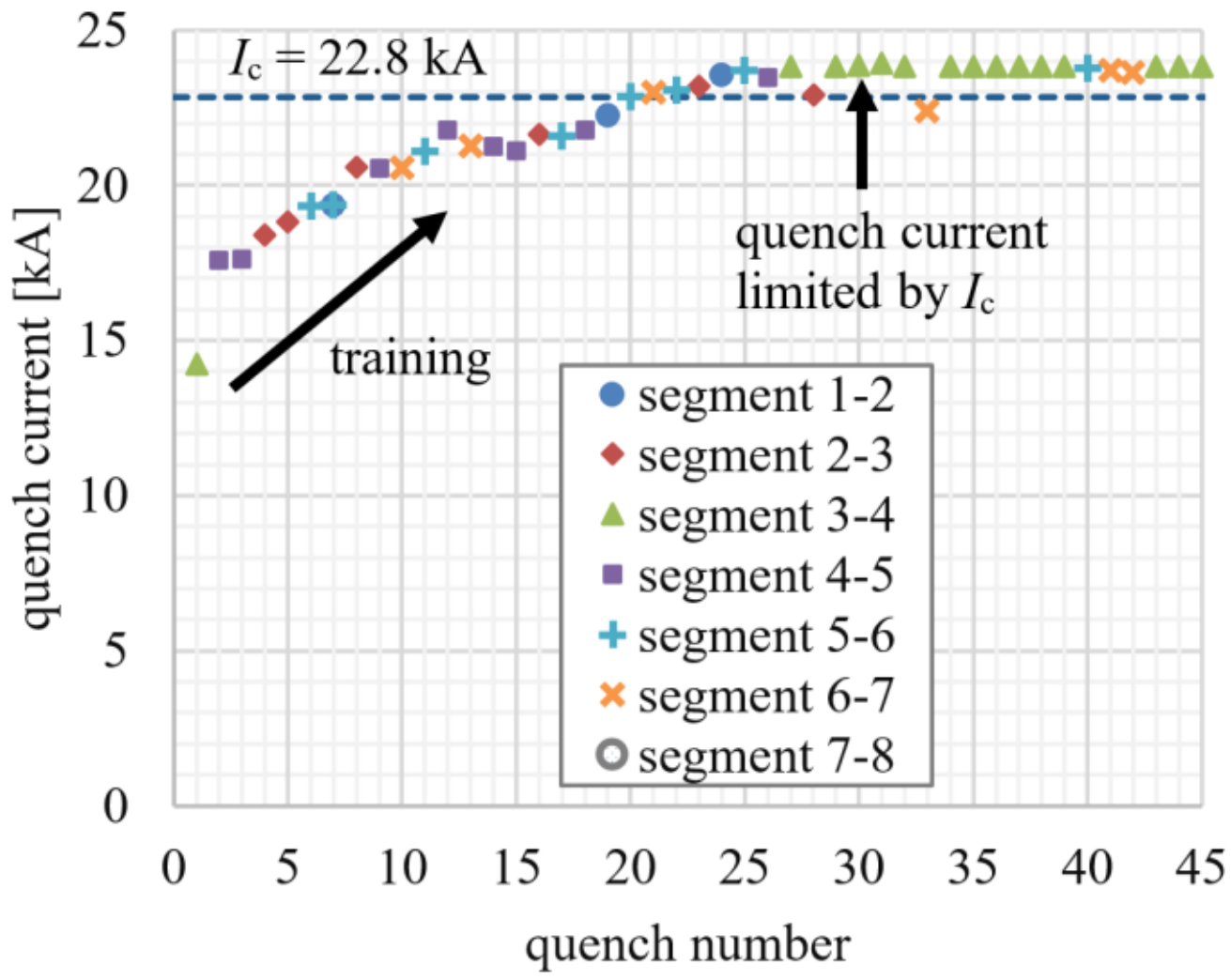

Fig. 2. Training curve of the BOX sample impregnated with Araldite MY750. The symbols correspond to quenches starting in different segments.

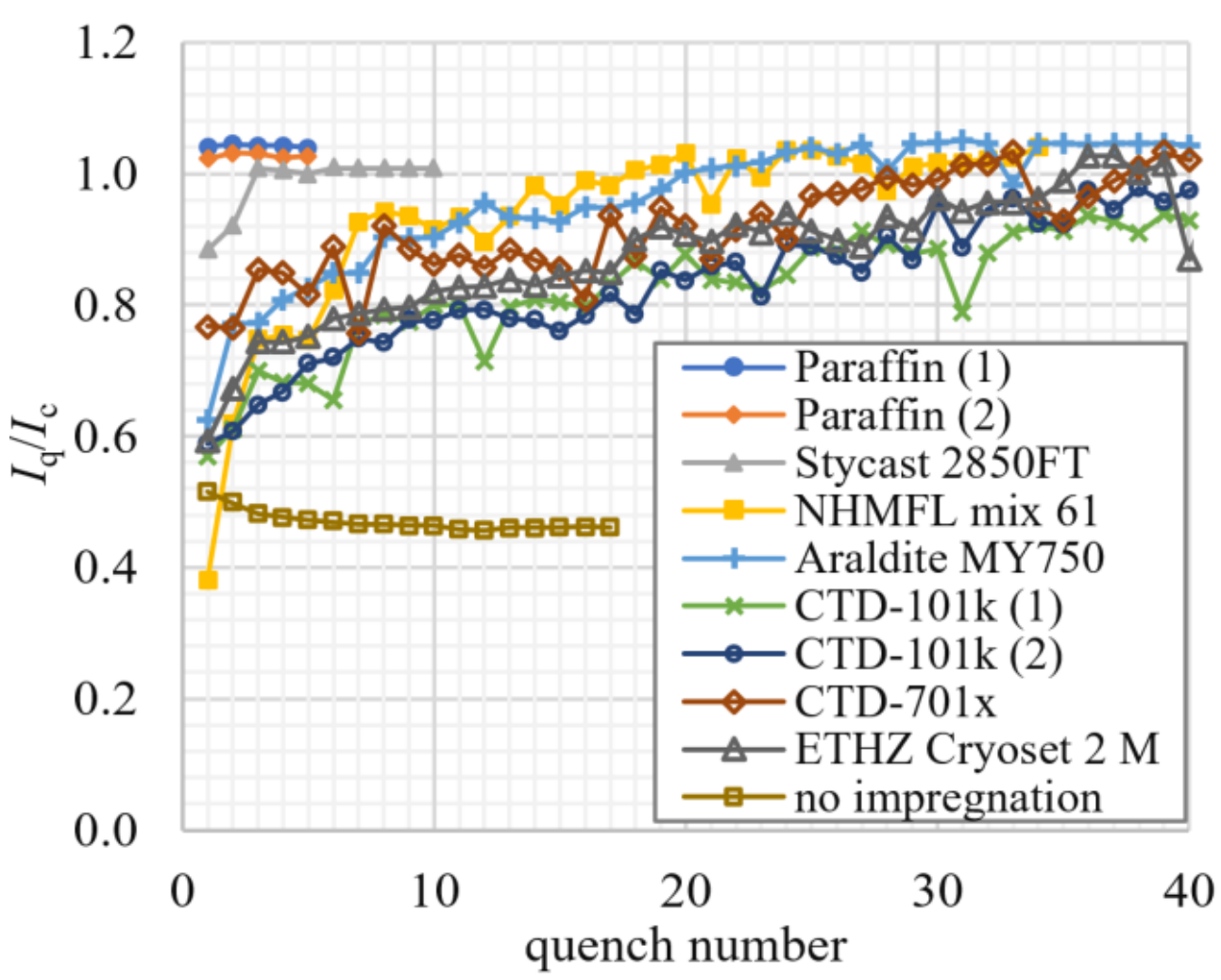

Fig. 3. Training curves for different resins. The quench currents $I_q$ of impregnated samples are normalized to their critical current $I_c$. The quench currents of the heavily damaged sample without impregnation are normalized to the average $I_c$ of other samples of 23.6 kA.

Figure 4 compares all samples impregnated with CTD-101K. Two unmodified samples were prepared in the same way (CTD-101K (1/2)). The similar quench currents of 57% and 59% of the critical current and overall similarity of the training curves demonstrate that the experiment is repeatable. The sample filled with $Al_2O_3$ had a higher initial quench current of 71% of the critical current compared to 52% to 59% for other samples impregnated with CTD-101K. Cross-sections of the alumina-filled sample revealed some voids, which may have limited the training performance. Therefore we plan a new test with a more properly filled sample. No increase in initial quench current was observed for the samples with double glass sleeves and weakened resin.

The initial quench currents for all impregnated samples are listed in Figure 5. For an objective comparison of training behavior between different samples, we define as criterion the number of training quenches before 80% of the critical current is reached. The Lorentz force at this current ranges from 171 kN/m to 188 kN/m depending on the critical current of the sample. This corresponds to an average shear stress of 8.7 MPa to 9.6 MPa if all force would be transferred to the impregnant through the wide surface of the cable. The number of quenches before reaching 80% of the critical current is shown in Figure 6. It is noted that the six samples with the highest initial quench current also reach 80% of the critical current in the same order.

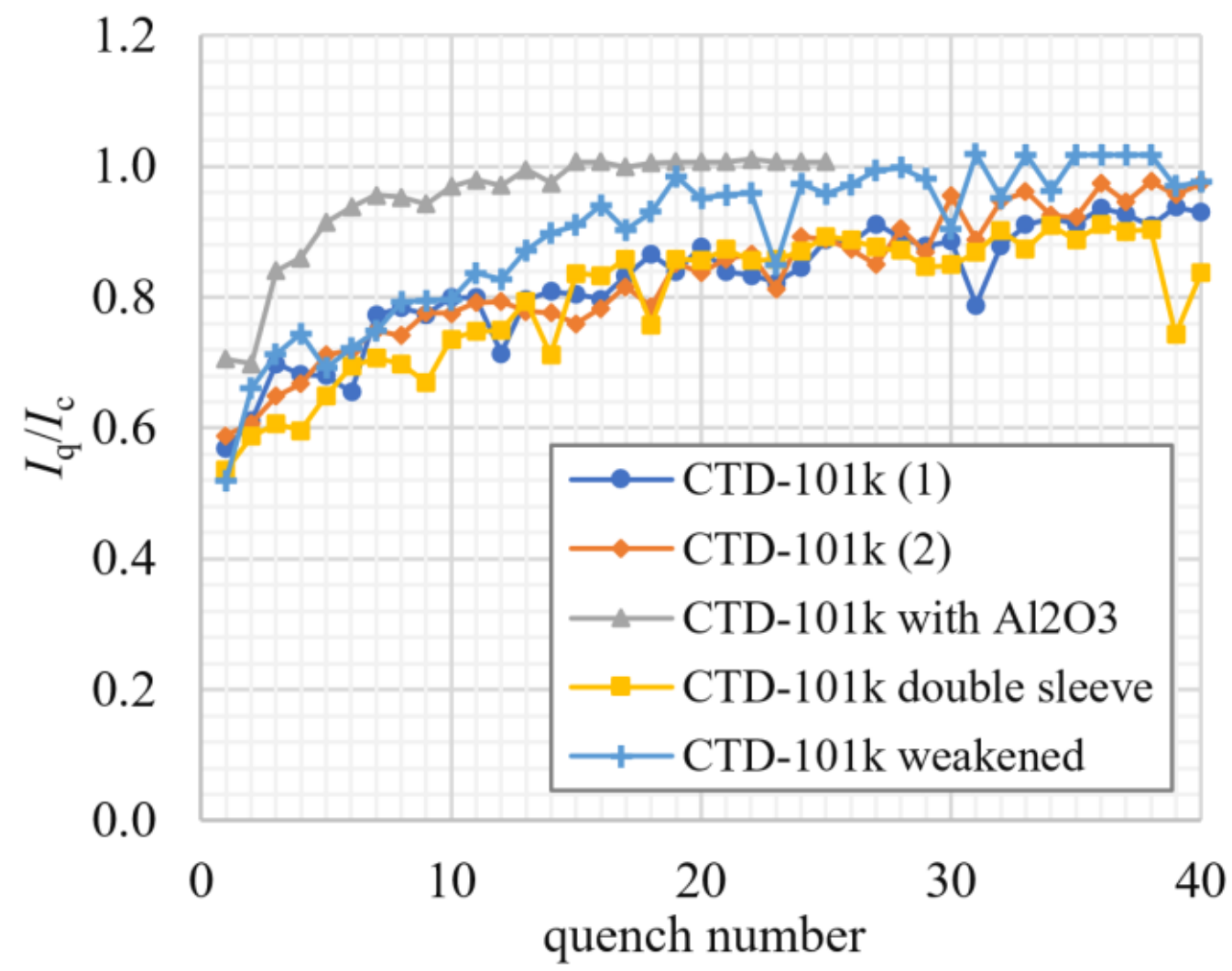

Fig. 4. Training curves for samples impregnated with CTD-101K.

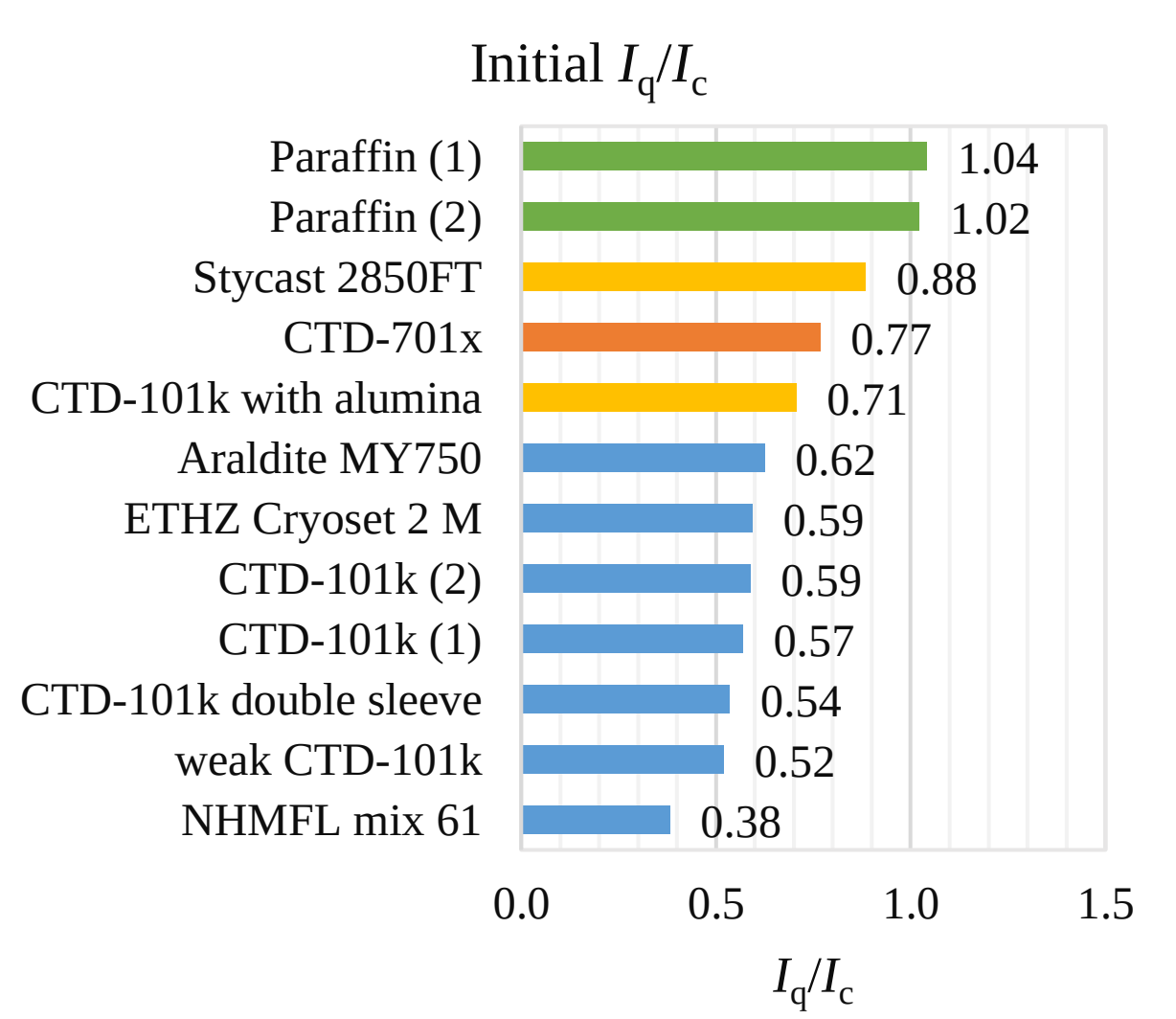

Fig. 5. Initial quench current normalized to the critical current for impregnated samples. Unfilled epoxies are shown in blue, alumina-filled epoxies in orange, polyolefin resin in red, and paraffin in green.

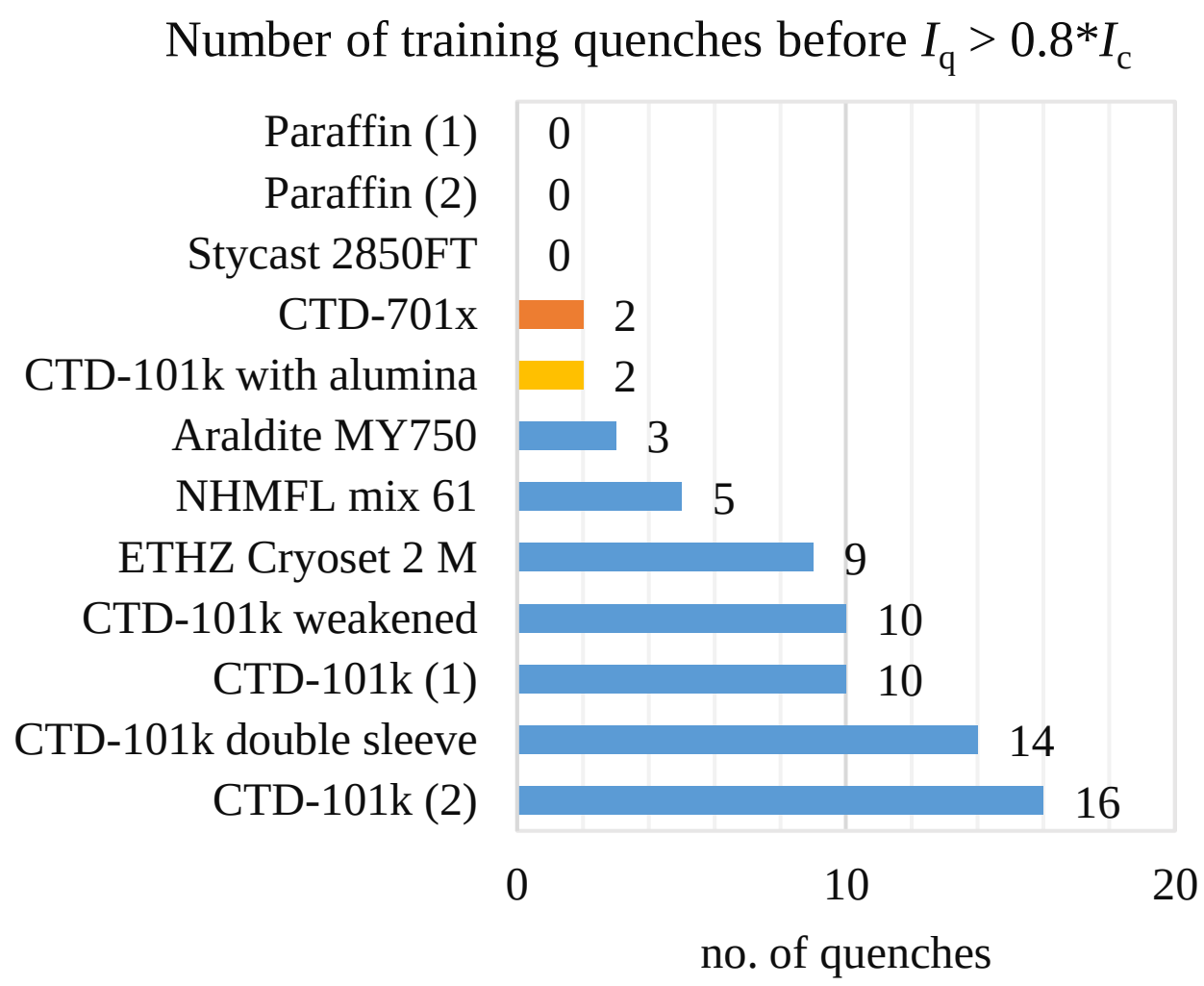

Fig. 6. Number of training quenches before 80% of the critical current is reached.

## B. *Critical currents*

The critical current was measured at applied magnetic fields of 7.5 T and 10 T and is listed in Table III. Some samples were not stable enough for a measurement at 7.5 T. For these samples, the critical current was estimated from the value at 10 T and the $I_c$ ratio between 7.5 T and 10 T for other samples of 1.44. The critical currents at 7.5 T range from 22.8 kA to 25.1 kA with an average of 23.6 kA. This excludes the degraded cable without impregnation, which had a critical current of only 6.8 kA.

TABLE III
CRITICAL CURRENT MEASUREMENTS

| Sample/impregnation | $B_a$ = 7.5 T | $B_a$ = 10 T |
|---|---|---|
| Paraffin (1) | 23.6 kA | 16.5 kA |
| Paraffin (2) | 24.0 kA | 16.7 kA |
| Stycast 2850FT/23LV | 25.1 kA | 17.2 kA |
| NHMFL mix 61 | 23.3 kA | 16.4 kA |
| Araldite MY750 | 22.8 kA | 15.9 kA |
| CTD-101K (1) | 22.8 kA* | 15.9 kA |
| CTD-101K (2) | 23.0 kA* | 16.0 kA |
| CTD-101K with $Al_2O_3$ | 23.9 kA* | 16.6 kA |
| CTD-101K double sleeve | 23.6 kA** | - |
| CTD-101K weakened | 24.3 kA* | 16.9 kA |
| CTD-701x | 23.6 kA** | - |
| ETHZ Cryoset 2 M | 23.0 kA | 16.0 kA |
| No impregnation | 6.8 kA*** | 4.4 kA*** |
| Average | 23.6 kA | 16.4 kA |

* Critical current at 7.5 T estimated from value at 10 T.
** Critical current not measured, average value of other samples used.
*** The sample without impregnation was heavily degraded and its $I_c$ data was excluded from the average.

# IV. CONCLUSION

In addition to paraffin [8], favorable training behavior was observed in Rutherford cables impregnated with alumina-filled epoxies. A sample impregnated with alumina-filled epoxy resin Stycast 2850FT had an initial quench current at 88% of the critical current, which is higher than in all other resin-impregnated samples. Another sample was filled with $Al_2O_3$ particles by sedimentation followed by impregnation with CTD-101K epoxy resin. This sample had a higher initial quench current (71% of $I_c$) compared to samples with unfilled CTD-101K (57% to 59%). The sample also reached 80% of $I_c$ after 2 quenches compared to 10 to 16 for the unfilled samples, despite the imperfect filling. It should be noted that the filled and unfilled samples have different insulation systems, because the glass braid is incompatible with filled resins. Nevertheless, the results indicate that the use of alumina filler may help to reduce training performance in $Nb_3Sn$ accelerator magnets. The physics behind this improvement are still to be investigated.

# ACKNOWLEDGMENT

We thank Ian Pong and his colleagues at LBNL for providing the Rutherford cable. This work was performed under the auspices and with support from the Swiss Accelerator Research and Technology (CHART) program (www.chart.ch).